\documentclass[12pt,thmsa]{article}
\usepackage{amssymb}
\usepackage{sw20lart}

\input tcilatex
\QQQ{Language}{
American English
}

\begin{document}

\author{Juan Antonio Valiente Kroon\thanks{%
E-mail: j.a.valiente@qmw.ac.uk} \\
%EndAName
School of Mathematical Sciences,\\
Queen Mary \& Westfield College,\\
Mile End Road, London E1 4NS.}
\title{Conserved Quantities for Polyhomogeneous Space-Times.}
\date{\TeXButton{today}{\today
}}
\maketitle

\begin{abstract}
The existence of conserved quantities with a structure similar to the
Newman-Penrose quantities in a polyhomogeneous space-time is addressed. The
most general form for the initial data formally consistent with the
polyhomogeneous setting is found. The subsequent study is done for those
polyhomogeneous space-times where the leading term of the shear $\sigma $
contains no logarithmic terms. It is found that for these space-times the
original NP quantities cease to be constants, but it is still possible to
construct a set of other 10 quantities that are constant. From these
quantities it is possible to obtain as a particular case a conserved
quantity found by Chru\'{s}ciel et al.
\end{abstract}

\section{Introduction.}

In a classical article by Newman \& Penrose \cite{n-p65}(see also \cite
{n-p68}, and \cite{ix} for a different treatment using the Bondi-Sachs
metric) the existence of ten conserved quantities (NP quantities) for the
gravitational field was established in a setting that assumed asymptotic
flatness. Recently Chru\'{s}ciel, MacCallum \& Singleton \cite{xiv} have
considered ``polyhomogeneous'' space-times as an adequate setting for
discussing the Bondi-Sachs characteristic initial value problem, showing
that polyhomogeneity at null infinity ($\frak{I}$ ) is formally consistent
with the Einstein field equations. They found that for the axisymmetric case
and assuming what they call the ``minimal sequence'' for the polyhomogeneity
of the initial data there is a conserved quantity that resembles those of
Newman \& Penrose. However, the general case (no symmetries, and arbitrary
polyhomogeneity) was left as an open question. The objective of this article
is to discuss the existence of the analogous of the Newman-Penrose conserved
quantities for a broad family of polyhomogeneous space-times.

The organization of this article is as follows: in section 2 the coordinate
system and the null tetrad used in this work are defined. In section 3 some
remarks concerning the asymptotic characteristic initial value problem are
made; the most general form for the logarithmic terms of the Weyl tensor
formally consistent with the field equations is investigated. Once the
general form is found, we restrict our study to those polyhomogeneous
space-times where the leading term of the shear ($\sigma $) contains no
logarithms. A possible physical interpretation of these space-times is
discussed. In section 4 the existence of conserved quantities similar in
structure to the NP quantities is discussed. It is found that for the
space-times under study the original NP quantities cease to be constant, but
nevertheless it is possible to construct some other constants (10 in
general). The relation with a constant quantity found by Chrusciel et al.
for the axisymmetric case is shown. Finally in section 5 some concluding
remarks are made. There are also 2 appendices. In appendix A there are two
tables showing the polyhomogeneous behaviour of the components of the Weyl
tensor, spin coefficients and tetrad functions for the general
polyhomogeneous case, and the case where $\sigma $ contains no logarithmic
terms. Some relations obtained from the expansions of the Bianchi identities
are also listed. In appendix B some useful properties of the differential
operator $\eth $ are listed for quick reference.

\section{Coordinate system, null tetrad \& the Newman-Penrose formalism.}

When studying the asymptotics of the Einstein field equations, a well suited
coordinate system can be constructed. The construction shown here is fairly
standard and a good reference is J. Stewart's book \cite{stewart}. As
mentioned in the introduction, the NP formalism will be used, and we will
refer to the field equations, commutation relations and Bianchi identities
as they are labeled in reference \cite{stewart}.

The first step is to introduce a family of null hypersurfaces in the
Riemannian manifold. We can use a parameter $u\,$ to label the hypersurfaces
of the family by $u=const.$; this defines a scalar field. As usual, the
first tetrad vector will be chosen to be orthogonal to the null
hypersurfaces,

\begin{equation}
\mathbf{l}=\mathbf{d}u.
\end{equation}
Since the hypersurfaces are null, the vectors $l^\mu $ will be tangent to a
family of curves on the hypersurfaces $\gamma _u$(the generators of the
hypersurfaces). These curves are null geodesics,

\begin{equation}
\mathbf{\nabla }\overrightarrow{_l}\overrightarrow{l}=0.
\end{equation}

On each generator $\gamma _u$ we can choose an arbitrary affine parameter $r$%
. Each null hypersurface $u=const$ will intersect $\frak{I}$ in a cut $S_u$.
On any cut it is possible to introduce arbitrary coordinates $x^i$ ($i=2,3$%
). The coordinates $x^i$ are propagated by demanding $x^i=const.$ on the
generators of $\frak{I}$ and on the generators of the null hypersufaces $%
\gamma _u$. In this way we have defined a coordinate system $%
(x^0,x^1,x^2,x^3)=(u,r,x^2,x^3)$ on a neighborhood of $\frak{I}$. There is
still some freedom remaining in the coordinate system just defined,
consisting of: a relabeling of the null hypersurfaces; a different choice of
the coordinates $x^i$ on the cuts $S_u$; and a freedom of the scaling and
origin of $r$ (remember it is an affine parameter).

The freedom on the scaling of $r$ can be used to set

\begin{equation}
\overrightarrow{l}=\frac \partial {\partial r}\text{ .}
\end{equation}
Together with $\overrightarrow{l}$ it is possible to define another null
vector $\overrightarrow{n}$ normalized by $\overrightarrow{l}\cdot 
\overrightarrow{n}=1$. The surfaces $u=const.$ and $r=const.$ will be
denoted by $S_{u,r}$. The last two vectors of the tetrad $\overrightarrow{m}$
and $\overrightarrow{\overline{m}}$ are chosen to span $T(S_{u,r})$ (surface
forming). The freedom left in this choice is a boost and a spin.

From the construction we get that the components of the tetrad can be
written as,

\begin{eqnarray}
l^\mu &=&\delta _1^\mu , \\
n^\mu &=&\delta _0^\mu +Q\delta _1^\mu +C^j\delta _j^\mu , \\
m^\mu &=&\xi ^i\delta _i^\mu ,
\end{eqnarray}
where $Q$, $C^i$, $\xi ^i$ ($i=2,3$) are complex functions of the
coordinates. Because $\overrightarrow{m}$ \& $\overrightarrow{\overline{m}}$
span $T(S_{u,r})$ then,

\begin{eqnarray}
\xi ^i\xi _i &=&0\text{,}  \nonumber \\
\xi ^i\overline{\xi }_i &=&-1\text{ (normalization).}
\end{eqnarray}

Applying the commutators to $x^i$ we get that:

\begin{eqnarray}
\kappa &=&\epsilon =0,  \nonumber \\
\tau &=&\overline{\alpha }+\beta ,  \nonumber \\
\mu &=&\overline{\mu },  \nonumber \\
\rho &=&\overline{\rho }.
\end{eqnarray}

\section{Initial data for the asymptotic characteristic initial value
problem.}

The NP quantities arise from considering the asymptotic characteristic
initial value problem for the vacuum Einstein field equations \cite{n-p65}, 
\cite{n-p68}. The coordinate system constructed before is well suited for
the study of this problem because with it the NP field equations form a
hierarchy of differential equations that is reasonably easy to solve under
certain assumptions. So, it will be necessary to make some remarks on the
kind of initial data required to have the problem well posed.

K\'{a}nn\'{a}r \cite{Kannar96} has proved an existence and uniqueness
theorem for the asymptotic characteristic initial value problem for the
vacuum field equations in the case of $C^\infty $ initial data. The initial
data is given on an incoming null hypersurface $\mathcal{N}$ and on part of
the past null infinity $\frak{I}^{-}$ (or of the future null infinity $\frak{%
I}^{+}$) which intersect in a two-dimensional space-like surface $\mathcal{Z}
$ diffeomorphic to $S^2$.

The $C^\infty $ initial data that assures existence of a unique solution in
a neighborhood of $\mathcal{Z}=\frak{I}^{-}\cap \mathcal{N}$ (or $\frak{I}%
^{+}\cap \mathcal{N}$) is given by:

\begin{eqnarray}
&&\widehat{\sigma }\text{ on }\frak{I}^{-}\text{ ( or }\frak{I}^{+}\text{),}
\nonumber \\
&&\widehat{\Psi }_0\text{ on }\mathcal{N}\text{, }  \nonumber \\
\text{and }\widehat{\Psi }_1\text{, }\widehat{\Psi }_2+\widehat{\overline{%
\Psi }}_2 &=&2\func{Re}\text{ }\Psi _2\text{, }\xi ^i\text{ on }\mathcal{Z}%
\text{,}
\end{eqnarray}
where the hatted quantities belong to the unphysical space-time. This is not
the only possible choice of initial data that fits K\'{a}nn\'{a}r's theorem,
but it is the one that is generally used when trying to solve the field
equations with asymptotic expansions. We have to point out here that
K\'{a}nn\'{a}r's theorem works only for $C^\infty $ initial data. So far, we
still do not have an existence/uniqueness theorem for polyhomogeneous
initial data; however Winicour \cite{winicour} has constructed some
space-times that possess a logarithmic behaviour similar to the one
discussed in this article.

In the Bondi-Sachs treatment the initial data on $\mathcal{N}$ is contained
in the metric functions $\gamma $ and $\delta $. So the information in $%
\gamma $ and $\delta $ is coded in $\Psi _0$ when working in the NP
formalism. In reference \cite{xiv} it was proved that if $\gamma $ and $%
\delta $ are polyhomogeneous functions of $r$, then the remaining metric
functions that are obtained from the initial data and the Einstein equations
are also polyhomogeneous. Due to the physical equivalence of the Bondi and
the Newman-Penrose treatment we have that if $\Psi _0$ is polyhomogeneous in 
$r$ then the NP spin coefficients that are calculated from this initial data
will be polyhomogeneous as well as the remaining components of the Weyl
tensor and the tetrad functions.

It can also be shown \cite{xiv} that given a sequence $\{N_i\}_{i=0}^\infty $
that defines the form of a certain polyhomogeneous function $f$ through

\begin{equation}
f=\sum_{i=1}^\infty \sum_{j=0}^{N_i}f_{ij}r^{-i}\ln ^jr,
\end{equation}
where the $f_{ij}$ are some functions of $(u,x^i)$, then there exists
another sequence $\{\widetilde{N}_i\}_{i=0}^\infty $ satisfying $N_i\leq 
\widetilde{N}_i$ which defines the form of the initial data $\gamma $ \& $%
\delta $ that is compatible with the field equations. This means that the
polyhomogeneous form of $\gamma $ \& $\delta $, or $\Psi _0$ in our case, is
not arbitrary, but there are heavy mathematical restrictions on the form of
the initial data. In particular we cannot put arbitrary powers of $\ln $ in
the leading terms of the initial data.

Because of the asymptotic character of our analysis we will not be concerned
with questions of convergence. We can always truncate the series to an
adequate order $M$ in the powers of $1/r$. For our purposes $M=6$ will be
enough.

Here arises the question of which is the most general form of $\Psi _0$ that
is compatible with the field equations. To answer this question we will
require some preliminary results.

Let, 
\begin{equation}
g=\sum_{i=1}^6g_i(z)r^{-i}+...\text{ },
\end{equation}
\begin{equation}
h=\sum_{i=1}^6h_i(z)r^{-i}+...\text{ },
\end{equation}
where $g_i(z)$ and $h_i(z)$ are polynomials in $z=\ln r$ be two
polyhomogeneous functions. It can be shown that, 
\begin{equation}
hg=\sum_{i=2}^6\text{ }\sum_{k=1}^{i-1}h_k(z)g_{i-k}(z)r^{-i}+(\text{terms
with higher powers of }1/r)\text{,}
\end{equation}
and that, 
\begin{eqnarray}
\frac \partial {\partial r}h &=&\sum_{i=2}^6\left( h_{i-1}^{\prime
}(z)-\left( i-1\right) h_{i-1}(z)\right) r^{-i}  \nonumber \\
&&+(\text{terms with higher powers of }1/r)\text{,}
\end{eqnarray}
where the apostrophe $^{\prime }$ denotes differentiation with respect to $z$%
.

Now, the most general form for $\Psi _0$ in the polyhomogeneous setting is 
\begin{equation}
\Psi _0=\sum_{i=1}^6\Psi _0^i(z)r^{-i}+...\text{ .}
\end{equation}
Let $N$ be the maximum of the degrees of the polynomials

\begin{equation}
\Psi _0^i(z)=\Psi _0^{i,N_i}z^{N_i}+\Psi _0^{i,N_i-1}z^{N_i-1}+...+\Psi
_0^{i,0},
\end{equation}
$i=1...6$, where $\Psi _0^{i,j}$ depend only on $(u,\theta ,\varphi )$.

To determine the restrictions on $\Psi _0$ we will use the first two field
equations in the NP hierarchy, namely equations (a) \& (b) of \cite{stewart},

\begin{equation}
D\rho =\rho ^2+\sigma \overline{\sigma },  \label{aa}
\end{equation}
\begin{equation}
D\sigma =2\rho \sigma +\Psi _0.  \label{bb}
\end{equation}
We will take $\rho $ \& $\sigma $ to be in principle of the same form as $%
\Psi _0$.

\begin{equation}
\rho =\sum_{i=1}^6\rho _ir^{-i}+...\text{,}
\end{equation}

\begin{equation}
\sigma =\sum_{i=1}^6\sigma _ir^{-i}+...\text{ .}
\end{equation}
Substitution into (a) \& (b) yields the following useful relations,

\begin{equation}
\rho _{i-1}^{\prime }-\left( i-1\right) \rho _{i-1}=\sum_{k=1}^{i-1}\left(
\rho _k\rho _{i-k}+\sigma _k\overline{\sigma }_{i-k}\right) ,  \label{A}
\end{equation}
and 
\begin{equation}
\sigma _{i-1}^{\prime }-\left( i-1\right) \sigma
_{i-1}=2\sum_{k=1}^{i-1}\rho _k\sigma _{i-k}+\Psi _0^i,  \label{B}
\end{equation}
for $i=2,3...$ $.$ Using these formulae, after a long but not difficult
analysis it is possible to show that the most general form for a
polyhomogeneous $\Psi _0$ formally consistent with (\ref{aa}) and (\ref{bb})
is

\begin{equation}
\Psi _0=\Psi _0^3[\left\| N/2\right\| -1]r^{-3}+\Psi _0^4[N-1]r^{-4}+\Psi
_0^5[N]r^{-5}+\Psi _0^6[N]r^{-6}\ln ^ir+...,  \label{psi0}
\end{equation}
where the quantities in square brackets are the degrees of the different
polynomials in $z=\ln r$. Here $\left\| N/2\right\| $ denotes the integer
part of $N/2$ (i.e. $\left\| 1/2\right\| =0,$ $\left\| 1\right\| =1,$ $%
\left\| 3/2\right\| =1...$).

As a by-product we also obtain the form for $\rho $, and $\sigma $,

\begin{equation}
\rho =-r^{-1}+\rho _3[N-1]r^{-3}+...  \label{rho}
\end{equation}

\begin{equation}
\sigma =\sigma _2[\left\| N/2\right\| ]r^{-2}+\sigma _3[N-1]r^{-3}+...\text{
.}  \label{sigma}
\end{equation}
Using the same technique we can obtain the leading terms for the remaining
spin coefficients, components of the Weyl tensor and tetrad functions. The
results are listed as a table in Appendix A. From that table we read

\begin{equation}
\Psi _1=\Psi _1^3[\left\| N/2\right\| -1]r^{-3}+\Psi _1^4[N]r^{-4}+...,
\label{psi1}
\end{equation}

\begin{equation}
\Psi _2=\Psi _2^3[\left\| N/2\right\| ]r^{-3}+\Psi _2^4[N]r^{-4}+...,
\label{psi2}
\end{equation}

\begin{equation}
\Psi _3=\Psi _3^2[0]r^{-2}+\Psi _2^3[\left\| N/2\right\| ]r^{-3}+...,
\label{psi3}
\end{equation}

\begin{equation}
\Psi _4=\Psi _4^1[0]r^{-1}+\Psi _4^2[0]r^{-2}+...\text{ .}  \label{psi4}
\end{equation}
Note that (\ref{psi0}) guarantees that $\Psi _n\rightarrow 0$ when $%
r\rightarrow \infty $ in agreement with the results of Couch \& Torrence 
\cite{ct72}, and the modified peeling behaviour of the Riemann tensor of
Chru\'{s}ciel et al. \cite{xiv}.

Note that in this setting the peeling theorem is no longer valid. The
introduction of coefficients with lower powers in $1/r$ ($\Psi _0^3$ and $%
\Psi _0^4$) translates into stronger incoming radiation fields than those
allowed by the so called \emph{outgoing radiation condition}. It was
believed that this condition ensured the non existence of incoming radiation
of infinite duration. In fact it can be showed that the $\Psi _0^4$
coefficient gives rise to the $1/r^2$ missing term in the original
Bondi-Sachs treatment. From the way these coefficients relate with the shear 
$\sigma $ (see the appendix A, and in particular equations (\ref{shear1})
and (\ref{shear2})) we can think of them as a contribution to the shear from
the incoming field. For a general polyhomogeneous space-time this shear will
dominate over the shear coming from the news function $\stackrel{.}{\sigma }%
_{2,0}$(i.e. shear associated with the outgoing radiation), because $%
r^{-2}\ln ^kr$ ($k>0$) falls to $0$ in a slower way than $r^{-2}$ when $%
r\rightarrow \infty $. Of course this interpretation is not completely
rigorous, and may be regarded only as a guideline.

This picture is somehow disturbing from the physical point of view, because
in the study of isolated gravitationally radiating systems we are mainly
interested in the outgoing radiation; so we would not like not to have its
effects overshadowed by the incoming radiation. We will demand that $%
\widehat{\sigma }$ is finite on $\frak{I}$, this is $\sigma _2=\sigma
_{2,0}(\theta ,\varphi )$ to be a polynomial of degree $0$ in $z=\ln r$ in
order to obtain space-times where the main origin of the shear is the
outgoing radiation. This fixes the form of the components of the Weyl tensor
to be:

\begin{equation}
\Psi _0=\Psi _0^4[N-1]r^{-4}+\Psi _0^5[N]r^{-5}+\Psi _0^6[N]r^{-6}\ln
^ir+...,  \label{PSI0}
\end{equation}

\begin{equation}
\Psi _1=\Psi _1^4[N]r^{-4}+...,  \label{PSI1}
\end{equation}

\begin{equation}
\Psi _2=\Psi _2^3[0]r^{-3}+\Psi _2^4[N]r^{-4}+...,  \label{PSI2}
\end{equation}

\begin{equation}
\Psi _3=\Psi _3^2[0]r^{-2}+\Psi _2^3[0]r^{-3}+...,  \label{PSI3}
\end{equation}

\begin{equation}
\Psi _4=\Psi _4^1[0]r^{-1}+\Psi _4^2[0]r^{-2}+...\text{ .}  \label{PSI4}
\end{equation}

We observe from the formulae (\ref{Bb1}), (\ref{Bd2}), and (\ref{Bf1}) that $%
\Psi _0^{4,N-1}$, $\Psi _1^{4,N}$, and $\Psi _2^{4,N}$ are constants of
motion (i.e their $u-$derivatives are zero). This is in agreement with the
statement of proposition 2.2 of \cite{xiv}.

\section{Conserved Quantities.}

As it is well known, Bianchi identities for vacuum in the NP formalism split
in two groups, those that contain $D$ and $\delta $ derivatives (Ba, Bc, Be,
Bg according to Stewart's labeling), and those with $\Delta $ and $\overline{%
\delta }$ derivatives (Bb, Bd, Bf \& Bh). The first group is used to
calculate the coefficients for the leading terms of the components of the
Weyl tensor on the initial hypersurface $\mathcal{N}$, while the $u$
dependence comes from the second set because $\Delta =n^\alpha \nabla
_\alpha =\frac \partial {\partial u}+Q\frac \partial {\partial r}+C^\alpha
\frac \partial {\partial x^\alpha }$. So, the Bianchi identities with $D$ \& 
$\overline{\delta }$ derivatives are constraint equations on the initial
data on the initial hypersurface and on $\frak{I}^{-}$ (or $\frak{I}^{+}$),
while the identities with $\Delta $ \& $\delta $ are the evolution equations
for the components of the Weyl tensor. Hence it is reasonable that if we
want to find conserved quantities formed from the initial data we have to
dig into the expansions for the Bianchi identity that contains $\Psi _0$ and 
$\Delta $ derivatives, namely (Bb). In what follows we will be integrating
over the unit sphere $S^2$, so it will be useful to rewrite the $\delta $
derivatives in terms of the eth operator $\eth $ and its conjugate. For this
purpose the relations

\begin{eqnarray}
\eth \eta &=&\delta \eta +s(\overline{\alpha }-\beta )\eta , \\
\overline{\eth }\eta &=&\overline{\delta }\eta -s(\alpha -\overline{\beta }%
)\eta \text{,}
\end{eqnarray}
will be most useful ($\eta $ is a quantity of spin weight $s$).

Following Newman \& Penrose \cite{n-p68}, we consider the $r^{-6}$%
coefficients for the expansion of Bianchi identity (Bb)$.$ Directly from the
expansion (equation (\ref{Bb3}) of the appendix A) we get:

\begin{eqnarray}
&&\stackrel{.}{\Psi }_0^6-\frac 12\Psi _0^{5\prime }+2\Psi _0^5-\frac
12\left( \Psi _2^3+\overline{\Psi }_2^3\right) \Psi _0^{4\prime }+2\overline{%
\Psi }_2^3\Psi _0^4-\eth \Psi _1^5+\sigma _{2,0}\overline{\eth }\Psi _1^4 
\nonumber  \label{r6} \\
&=&\eth \overline{\sigma }_{2,0}\eth \Psi _0^4+\overline{\eth }\sigma _{2,0}%
\overline{\eth }\Psi _0^4-5\Psi _2^3\Psi _0^4-\eth ^2\sigma _{2,0}\Psi
_0^4-\sigma _{2,0}\stackrel{.}{\overline{\sigma }}_{2,0}\Psi _0^4  \nonumber
\label{r6} \\
&&+\Psi _2^3\Psi _0^4-4\overline{\eth }\sigma _{2,0}\Psi _1^4+3\sigma
_{2,0}\Psi _2^4\text{.}  \label{r6}
\end{eqnarray}
Dotted quantities are derivatives with respect to $u$. From (\ref{r6}) we
can read the coefficients for the different powers of $z$. In this
expression, only the terms $\stackrel{.}{\Psi }_0^6$, $2\Psi _0^5$, $\eth
\Psi _1^5$, $\sigma _{2,0}\overline{\eth }\Psi _1^4$, $\overline{\eth }%
\sigma _{2,0}\Psi _1^4$, and $\sigma _{2,0}\Psi _2^4$ are polynomials of
degree $N$ in $z=\ln r$ (see table 2). So the coefficient for $r^{-6}\ln ^Nr$
is,

\begin{equation}
\stackrel{.}{\Psi }_0^{6,N}+2\Psi _0^{5,N}=3\sigma _{2,0}\Psi _2^{4,N}-4%
\overline{\eth }\sigma _{2,0}\Psi _1^{4,N}+\eth \Psi _1^{5,N}-\sigma _{2,0}%
\overline{\eth }\Psi _1^{4,N}.  \label{Bb6N}
\end{equation}
From equations (\ref{Ba2}) and (\ref{Bc1}) we can deduce

\begin{equation}
\Psi _1^{5,N}=-\overline{\eth }\Psi _0^{5,N},
\end{equation}
\begin{equation}
\Psi _2^{4,N}=-\overline{\eth }\Psi _1^{4,N}.
\end{equation}
Substitution of the later two expressions into (\ref{Bb6N}) yields:

\begin{equation}
\stackrel{.}{\Psi }_0^{6,N}=\left( -4\sigma _{2,0}\overline{\eth }\Psi
_1^{4,N}-4\overline{\eth }\sigma _{2,0}\Psi _1^{4,N}\right) +\left( \eth 
\overline{\eth }\Psi _0^{5,N}-2\Psi _0^{5,N}\right) \text{.}
\end{equation}
The terms in the first parenthesis are clearly the $\eth $-derivative of a
product, while in the second parenthesis one we can use the commutation
relation for $\eth $ \& $\overline{\eth }$ (equation (\ref{eth5})) yielding:

\begin{equation}
\stackrel{.}{\Psi }_0^{6,N}=-4\overline{\eth }(\sigma _{2,0}\Psi _1^{4,N})-%
\overline{\eth }\eth \Psi _0^{5,N}.  \label{integrand}
\end{equation}

The first two terms of the right hand side of equation (\ref{integrand}) are
of the form $\overline{\eth }$ applied to a quantity of spin weight 3.
Multiplying the last two equations by $_2\overline{Y}_{l,m}$ and integrating
over $S^2$ it is possible to apply formula (\ref{eth7}) in the case $s=2$
and obtain

\begin{equation}
\stackrel{}{\stackrel{.}{\mathcal{Q}}_m=\int_{S^2}\stackrel{.}{\Psi }_0^{6,N}%
}(_2\overline{Y}_{2,m})d\omega =0
\end{equation}
for $m=-2,-1,0,1,2$. Hence

\begin{equation}
\mathcal{Q}_m=\int_{S^2}\Psi _0^{6,N}(_2\overline{Y}_{2,m})d\omega
\label{conserved}
\end{equation}
gives 10 real conserved quantities.

If we try to apply the latter argument to the equations associated to the
other powers of $\ln r$ (and in particular to $\ln ^0r,$ from which we can
construct the NP quantities) we will find that in principle there are
several new terms (coming mainly from $\Psi _0^4,$ see equation (\ref{r6}))
that cannot be written as the$\overline{\text{ }\eth }$ derivative of
something. The fact that $\Psi _0^4$ is a polynomial of degree $N-1$ in $\ln
r$ happens to be an essential ingredient to obtain the conservation law.

\subsection{The conserved quantity of Chru\'{s}ciel et al.}

Chru\'{s}ciel et al. found that for an axisymmetric polyhomogeneous
space-time which follows the ``minimal sequence'' the quantity

\begin{equation}
\mathcal{Q}_{XIV}=\int_{S^2}\gamma _{41}\sin ^2\theta d\omega ,
\end{equation}
is conserved. In our treatment, we recover the leading terms of the minimal
sequence by setting $N=1$. It is not hard to prove then that for the
axisymmetric case (when all the components of the Weyl tensor are real)and $%
N=1$ that

\begin{equation}
\gamma _{41}=-\frac 1{12}\Psi _0^{6,1}.
\end{equation}
Now, it can be shown that $\sin ^2\theta $ is proportional to the spin
spherical harmonic $_2\overline{Y}_{2,0}$ \cite{ix} so that from (\ref
{conserved}) we can recover $\mathcal{Q}_{XIV}$ by putting $m=0$.

\section{Conclusions.}

It has been proven that for a polyhomogeneous space-time in which the
leading term of the shear ($\sigma _2$) has no logarithmic terms there are
ten conserved quantities. These polyhomogeneous space-times can be
interpreted as those where the shear of the outgoing gravitational radiation
dominates over the shear of the incoming radiation.

The attempts to interpret the NP quantities have been so far inconclusive
(see for example \cite{n-p68}, \cite{g-g70}). It also has been suggested
that these quantities lack a physical meaning \cite{bardeen}. However most
authors agree that the NP quantities reflect in some way the structure of
the incoming radiation. This opinion is reinforced with the conserved
quantities $\mathcal{Q}_m$, because they possess an analogous structure, and
because the kind of space-times we are working with admit much stronger
incoming gravitational radiation than the asymptotically flat space-times
used by Newman and Penrose.

It has to be mentioned that we suspect that for a general polyhomogeneous
space-time the quantities $\mathcal{Q}_m$ cease to be constants, or at least
the arguments used here cannot be applied anymore due to the appearance of
several new terms that hardly could be rewritten as a $\overline{\eth }$
derivative. In more physical terms this could be rephrased by saying that
the $\mathcal{Q}_m$ are no longer conserved because the shear of the
incoming radiation dominates over that of the outgoing radiation. In a
similar way the NP quantities are not constants in a polyhomogeneous setting
because the incoming radiation is not of finite duration (see for example 
\cite{n-p68}, \cite{j-n65}). A deeper exploration of these ideas will be the
subject of future work.

\section{Acknowledgments.}

I am most grateful to Prof. M.A.H. MacCallum for suggesting this topic, for
several discussions and for careful revisions of previous versions of this
article. Thanks also to Dr. P.T. Chru\'{s}ciel and Dr. M.\ Mars for helpful
discussions. The author has a scholarship (110441/110491) from the Consejo
Nacional de Ciencia y Tecnolog\'{\i}a (CONACYT), Mexico.

\appendix 

\section{Polyhomogeneous expansions in the NP formalism.}

\subsection{The relationship between $\rho $, $\sigma $ \& $\Psi _0$ for a
general polyhomogeneous space-time.}

Beginning with equations (\ref{A}) and (\ref{B})

\begin{equation}
\rho _{i-1}^{\prime }-\left( i-1\right) \rho _{i-1}=\sum_{k=1}^{i-1}\left(
\rho _k\rho _{i-k}+\sigma _k\overline{\sigma }_{i-k}\right) ,
\end{equation}
\begin{equation}
\sigma _{i-1}^{\prime }-\left( i-1\right) \sigma
_{i-1}=2\sum_{k=1}^{i-1}\rho _k\sigma _{i-k}+\Psi _0^i,
\end{equation}
we find for $i=1$ that

\begin{equation}
\Psi _0^1=0\text{;}
\end{equation}
for $i=2,$

\begin{eqnarray}
\rho _1 &=&\rho _{1,0}=-1, \\
\sigma _1 &=&0, \\
\Psi _0^2 &=&0.
\end{eqnarray}
For $i=3$ we find that

\begin{eqnarray}
\rho _2^{\prime } &=&0, \\
\sigma _2^{\prime } &=&\Psi _0^3.
\end{eqnarray}
So $\rho _2$ is a polynomial of degree 0 in $\ln r$. Using the remaining
freedom in the definition of $r$ (recall that $r$ is an affine parameter) we
can redefine $r$ such that $\rho _2=0$ (see for example \cite{n-u62}). From
the other equation we read

\begin{eqnarray}
\Psi _0^{3,N} &=&0, \\
\Psi _0^{3,j} &=&(j+1)\sigma _{2,j+1}.  \label{shear1}
\end{eqnarray}
Hence we can think of $\Psi _0^{3,j}$ $j=0...N-1$ as a contribution to the
shear from the incoming radiation using Szekeres interpretation of the
components of the Weyl tensor ($\Psi _0$ can be regarded as an incoming
transverse wave)\cite{szekeres}.

For $i=4$ we obtain, 
\begin{equation}
\sigma _{2,N}=\sigma _{2,N-1}=...=\sigma _{2,\left\| N/2\right\| }=0,
\end{equation}

\begin{eqnarray}
\Psi _0^{3,N-1} &=&...=\Psi _0^{3,\left\| N/2\right\| -1}=0, \\
\rho _{3,N} &=&0, \\
-\rho _{3,j}+(j+1)\rho _{3,j+1} &=&\sum_{k=0}^j\sigma _{2,k}\overline{\sigma 
}_{2,j-k\text{,}}
\end{eqnarray}
\begin{equation}
\sigma _{3,N}=-\Psi _0^{4,N},
\end{equation}

\begin{equation}
-\sigma _{3,j}+(j+1)\sigma _{3,j+1}=\Psi _0^{4,j}\text{.}  \label{shear2}
\end{equation}
for $j=0...N-1$. Using the relations for $i=6$ it is possible to find
further bounds on $\sigma _3$ and $\Psi _0^4$.

\subsection{Orders of the leading terms for a general polyhomogeneous
space-time.}

Using a similar technique we can set bounds for the orders of the
``logarithmic polynomials'' of the leading terms for the remaining spin
coefficients, components of the Weyl tensor and tetrad functions. The
degrees of the polynomials are listed as a table. The - means that such a
term does not appear in the expansion.

\begin{equation}
\begin{tabular}{|c|c|c|c|c|c|c|c|}
\hline
& $1$ & $r^{-1}$ & $r^{-2}$ & $r^{-3}$ & $r^{-4}$ & $r^{-5}$ & $r^{-6}$ \\ 
\hline
$\Psi _0$ & - & - & - & $\left\| N/2\right\| -1$ & $N-1$ & $N$ & $N$ \\ 
\hline
$\Psi _1$ & - & - & - & $\left\| N/2\right\| -1$ & $N$ & ... & ... \\ \hline
$\Psi _2$ & - & - & - & $\left\| N/2\right\| $ & $N$ & ... & ... \\ \hline
$\Psi _3$ & - & - & $0$ & $\left\| N/2\right\| $ & ... & ... & ... \\ \hline
$\Psi _4$ & - & $0$ & $0$ & ... & ... & ... & ... \\ \hline
$\rho $ & - & $0$ & $0$ & $N-1$ & $N$ & ... & ... \\ \hline
$\sigma $ & - & - & $\left\| N/2\right\| $ & $N$ & $N$ & ... & ... \\ \hline
$\alpha $ & - & $0$ & $\left\| N/2\right\| +1$ & $N$ & ... & ... & ... \\ 
\hline
$\beta $ & - & $0$ & $\left\| N/2\right\| $ & $N$ & ... & ... & ... \\ \hline
$\tau $ & - & - & $\left\| N/2\right\| +1$ & $N$ & ... & ... & ... \\ \hline
$\pi $ & - & - & $\left\| N/2\right\| +1$ & $N$ & ... & ... & ... \\ \hline
$\gamma $ & - & $0$ & $\left\| N/2\right\| +1$ & ... & ... & ... & ... \\ 
\hline
$\lambda $ & - & $0$ & $\left\| N/2\right\| +1$ & ... & ... & ... & ... \\ 
\hline
$\mu $ & - & $0$ & $\left\| N/2\right\| +1$ & ... & ... & ... & ... \\ \hline
$v$ & - & $0$ & ... & ... & ... & ... & ... \\ \hline
$\xi ^\alpha $ & - & $0$ & $\left\| N/2\right\| $ & $N$ & ... & ... & ... \\ 
\hline
$Q$ & $0$ & $\left\| N/2\right\| +1$ & ... & ... & ... & ... & ... \\ \hline
$C^\alpha $ & - & - & $\left\| N/2\right\| +1$ & ... & ... & ... & ... \\ 
\hline
\end{tabular}
\tag{Table 1}
\end{equation}

\subsection{Orders for the leading terms for polyhomogeneous space-times
such that $\sigma _2=\sigma _{2,0}(\theta ,\varphi ).$}

The restriction $\sigma _2=\sigma _{2,0}(\theta ,\varphi )$ puts lower
bounds on the orders of the ``logarithmic polynomials''. These are listed in
table 2.

\begin{equation}
\begin{tabular}{|c|c|c|c|c|c|c|c|}
\hline
& $1$ & $r^{-1}$ & $r^{-2}$ & $r^{-3}$ & $r^{-4}$ & $r^{-5}$ & $r^{-6}$ \\ 
\hline
$\Psi _0$ & - & - & - & - & $N-1$ & $N$ & $N$ \\ \hline
$\Psi _1$ & - & - & - & - & $N$ & ... & ... \\ \hline
$\Psi _2$ & - & - & - & $0$ & $N$ & ... & ... \\ \hline
$\Psi _3$ & - & - & $0$ & $0$ & ... & ... & ... \\ \hline
$\Psi _4$ & - & $0$ & $0$ & $0$ & ... & ... & ... \\ \hline
$\rho $ & - & $0$ & $0$ & $0$ & $N-1$ & $N$ & ... \\ \hline
$\sigma $ & - & - & $0$ & $N-1$ & $N$ & ... & ... \\ \hline
$\alpha $ & - & $0$ & $0$ & $N$ & ... & ... & ... \\ \hline
$\beta $ & - & $0$ & $0$ & $N$ & ... & ... & ... \\ \hline
$\tau $ & - & - & $0$ & $N$ & ... & ... & ... \\ \hline
$\pi $ & - & - & $0$ & $N$ & ... & ... & ... \\ \hline
$\gamma $ & - & - & $0$ & ... & ... & ... & ... \\ \hline
$\lambda $ & - & $0$ & $0$ & ... & ... & ... & ... \\ \hline
$\mu $ & - & $0$ & $0$ & ... & ... & ... & ... \\ \hline
$v$ & - & $0$ & ... & ... & ... & ... & ... \\ \hline
$\xi ^\alpha $ & - & $0$ & $0$ & $N-1$ & ... & ... & ... \\ \hline
$Q$ & $0$ & $0$ & ... & ... & ... & ... & ... \\ \hline
$C^\alpha $ & - & - & $0$ & ... & ... & ... & ... \\ \hline
\end{tabular}
\tag{Table2}
\end{equation}
So that the leading terms are,

\begin{equation}
\Psi _0=\left( \sum_{k=0}^{N-1}\Psi _0^{4,k}\ln ^kr\right) r^{-4}+...\text{ ,%
}
\end{equation}

\begin{equation}
\Psi _1=\left( \sum_{k=0}^N\Psi _1^{4,k}\ln ^kr\right) r^{-4}+...\text{ ,}
\end{equation}

\begin{equation}
\Psi _2=\left( \Psi _2^{3,0}\right) r^{-3}+...\text{ ,}
\end{equation}

\begin{equation}
\Psi _3=-\eth \stackrel{.}{\overline{\sigma }}_{2,0}r^{-2}+...\text{ ,}
\end{equation}

\begin{equation}
\Psi _4=-\stackrel{..}{\overline{\sigma }}_{2,0}r^{-1}+...\text{ ,}
\end{equation}

\begin{equation}
\rho =-r^{-1}-(\sigma _{2,0}\overline{\sigma }_{2,0})r^{-3}+...\text{ ,}
\end{equation}

\begin{equation}
\sigma =\sigma _{2,0}r^{-2}+...\text{ ,}
\end{equation}

\begin{equation}
\alpha =\alpha _{1,0}r^{-1}+\left( \eth \overline{\sigma }_{2,0}+\overline{%
\alpha }_{1,0}\overline{\sigma }_{2,0}\right) r^{-2}+...\text{ ,}
\end{equation}

\begin{eqnarray}
\beta &=&-\overline{\alpha }_{1,0}r^{-1}-\alpha _{1,0}\sigma
_{2,0}r^{-2}+....\text{ ,}
\end{eqnarray}

\begin{equation}
\tau =\overline{\eth }\sigma _{2,0}r^{-2}+...\text{ , }
\end{equation}

\begin{equation}
\pi =\eth \overline{\sigma }_{2,0}r^{-2}+\text{ }...\text{ ,}
\end{equation}

\begin{eqnarray}
\gamma &=&-\left( \alpha _{1,0}\overline{\eth }\sigma _{2,0}-\overline{%
\alpha }_{1,0}\eth \overline{\sigma }_{2,0}+\frac 12\Psi _2^{3,0}\right)
r^{-2}+...\text{ ,}
\end{eqnarray}

\begin{equation}
\mu =-\frac 12r^{-1}+\left( \eth ^2\overline{\sigma }_{2,0}+\sigma _{2,0}%
\stackrel{.}{\overline{\sigma }}_{2,0}+\Psi _2^{3,0}\right) r^{-2}+...\text{
,}  \nonumber
\end{equation}

\begin{equation}
\lambda =\stackrel{.}{\overline{\sigma }}_{2,0}r^{-1}-\left( \overline{\eth }%
\eth \overline{\sigma }_{2,0}-\frac 12\overline{\sigma }_{2,0}\right)
r^{-2}+...\text{ ,}
\end{equation}

\begin{equation}
v=\frac 12\eth \left( \Psi _2^3+\overline{\Psi }_2^3\right) r^{-2}+...
\end{equation}

\begin{equation}
\text{ and }\kappa =\epsilon =0\text{,}
\end{equation}

\begin{equation}
\text{ where }\alpha _{1,0}=-\frac 1{2\sqrt{2}}\cot \theta .
\end{equation}

\subsection{The tetrad functions.}

The tetrad functions are calculated from the frame equations that are
obtained from applying the commutators of the different derivatives ($D$, $%
\Delta $, $\delta $, $\overline{\delta }$) to $u$, $r$, and to $x^i$. The
results are,

\begin{equation}
\xi ^\alpha =\xi _{1,0}^{\alpha \text{ }}r^{-1}-\sigma _{2,0}\overline{\xi }%
_{1,0}^\alpha r^{-2}+...\text{ ,}
\end{equation}

\begin{eqnarray}
C^i &=&-\left( \eth \overline{\sigma }_{2,0}\xi ^{0i}+\overline{\eth }\sigma
_{2,0}\overline{\xi }^{0i}\right) r^{-2}+...\text{ ,}
\end{eqnarray}

\begin{eqnarray}
Q &=&-\frac 12-\frac 12\left( \Psi _2^3+\overline{\Psi }_2^3\right)
r^{-1}+...\text{ ,}  \nonumber
\end{eqnarray}

where

\begin{eqnarray}
\xi ^{0\theta } &=&\frac 1{\sqrt{2}}, \\
\xi ^{0\varphi } &=&-\frac i{\sqrt{2}}\csc \theta \text{.}
\end{eqnarray}

\subsection{Relations obtained from the Bianchi identities when $\sigma
_2=\sigma _{2,0}(\theta ,\varphi )$.}

\subsubsection{(Ba).}

\begin{equation}
\Psi _1^{4^{\prime }}=\overline{\eth }\Psi _0^4,  \label{Ba1}
\end{equation}

\begin{equation}
\Psi _1^{5^{\prime }}-\Psi _1^5=\overline{\eth }\Psi _0^5+\left( \overline{%
\beta }_2-3\alpha _2\right) \Psi _0^4.  \label{Ba2}
\end{equation}

\subsubsection{(Bb).}

\begin{equation}
\stackrel{.}{\Psi }_0^{4,N-1}=0\text{ },  \label{Bb1}
\end{equation}

\begin{equation}
\stackrel{.}{\Psi }_0^{5,N}=\eth \Psi _1^{4,N},  \label{Bb2}
\end{equation}

\begin{eqnarray}
&&\stackrel{.}{\Psi }_0^6-\frac 12\Psi _0^{5\prime }+2\Psi _0^5-\frac
12\left( \Psi _2^3+\overline{\Psi }_2^3\right) \Psi _0^{4\prime }+2\overline{%
\Psi }_2^3\Psi _0^4-\eth \Psi _1^5+\sigma _{2,0}\overline{\eth }\Psi _1^4 
\nonumber  \label{r6} \\
&=&\eth \overline{\sigma }_{2,0}\eth \Psi _0^4+\overline{\eth }\sigma _{2,0}%
\overline{\eth }\Psi _0^4-5\Psi _2^3\Psi _0^4-\eth ^2\sigma _{2,0}\Psi
_0^4-\sigma _{2,0}\stackrel{.}{\overline{\sigma }}_{2,0}\Psi _0^4  \nonumber
\label{r6} \\
&&+\Psi _2^3\Psi _0^4-4\overline{\eth }\sigma _{2,0}\Psi _1^4+3\sigma
_{2,0}\Psi _2^4\text{.}  \label{Bb3}
\end{eqnarray}

\subsubsection{(Bc).}

\begin{equation}
\Psi _2^{4^{\prime }}-\Psi _2^4=\stackrel{.}{\overline{\sigma }}_{2,0}\Psi
_0^4+\overline{\eth }\Psi _1^4.  \label{Bc1}
\end{equation}

\subsubsection{(Bd).}

\begin{equation}
\stackrel{.}{\Psi }_1^{4,N}=0,  \label{Bd1}
\end{equation}

\begin{equation}
\stackrel{.}{\Psi }_1^{5,N}=\eth \Psi _2^{4,N}-\Psi _1^{4,N}.  \label{Bd2}
\end{equation}

\subsubsection{(Be).}

\begin{equation}
\Psi _3^{3,0}=-\overline{\eth }\Psi _2^{3,0}.  \label{Be1}
\end{equation}

\subsubsection{(Bf).}

\begin{equation}
\stackrel{.}{\Psi }_2^{4,N}=0.  \label{Bf1}
\end{equation}

\subsubsection{(Bg).}

\begin{equation}
\Psi _4^{2,0}=-\overline{\eth }\Psi _3^{2,0}.  \label{Bg1}
\end{equation}

\section{Properties of $\eth $ \& $\overline{\eth }.$}

The properties of the differential operator $\eth $ relevant for the present
article are presented here as a quick reference. For proofs and a more
extended discussion refer to \cite{n-p68}, \cite{BMS}, \cite{goldberg67}.

A quantity $\eta $ is said to have spin weight $s$ if under the tetrad
transformation

\begin{equation}
m^{\mu \prime }=e^{i\psi }m^\mu ,  \label{eth1}
\end{equation}
it transforms as 
\begin{equation}
\eta ^{\prime }=e^{si\psi }\eta \text{.}  \label{eth2}
\end{equation}
For example $\sigma $ has spin weight 2, and $\Psi _k$ has spin weight $2-k$%
. Let now $\eta $ be a quantity of spin weight $s$ defined on the sphere. We
define the operators $\eth $ and $\overline{\eth }$ by

\begin{equation}
\eth \eta =-\left( \sin \theta \right) ^s\left\{ \frac \partial {\partial
\theta }+\frac i{\sin \theta }\frac \partial {\partial \theta }\right\}
\left\{ \left( \sin \theta \right) ^{-s}\eta \right\} \text{,}  \label{eth3}
\end{equation}

\begin{equation}
\overline{\eth }\eta =\left( \sin \theta \right) ^{-s}\left\{ \frac \partial
{\partial \theta }-\frac i{\sin \theta }\frac \partial {\partial \theta
}\right\} \left\{ \left( \sin \theta \right) ^s\eta \right\} \text{.}
\label{eth4}
\end{equation}
$\eth \eta $ has spin weight $s+1$, and $\overline{\eth }\eta $ has spin
weight $s-1$. Their commutator is:

\begin{equation}
\left( \eth \overline{\eth }-\overline{\eth }\eth \right) \eta =2s\eta \text{%
.}  \label{eth5}
\end{equation}
The spin $s$ spherical harmonics are defined as ($s$ integer),

\begin{equation}
_sY_{l,m}=\left\{ 
\begin{tabular}{ll}
$\left[ \dfrac{\left( l-s\right) !}{\left( l+s\right) !}\right] ^{1/2}\eth
^sY_{l,m}$ & for $0\leq s\leq l$, \\ 
$\left( -1\right) ^s\left[ \dfrac{\left( l+s\right) !}{\left( l-s\right) !}%
\right] \overline{\eth }^{-s}Y_{l,m}$ & for $-l\leq s\leq 0$.
\end{tabular}
\right.  \label{eth6}
\end{equation}
And finally we have that if $\zeta $ is of spin weight $l+1$ then,

\begin{equation}
\int_{S^2}\left( _s\overline{Y}_{l,m}\right) \overline{\eth }^{l-s+1}\zeta
d\omega =0\text{.}  \label{eth7}
\end{equation}

\end{document}